\documentclass[sigconf, nonacm]{acmart}

\AtBeginDocument{%
  \providecommand\BibTeX{{%
    \normalfont B\kern-0.5em{\scshape i\kern-0.25em b}\kern-0.8em\TeX}}}

\usepackage{amsmath,amssymb,amsfonts}
\usepackage{algorithmic}
\usepackage{graphicx}
\usepackage{textcomp}
\usepackage{subcaption}
\usepackage{xcolor}
\usepackage{multirow}
\usepackage{soul}

\renewcommand\footnotetextcopyrightpermission[1]{}
\begin{document}

\title{CRAM-ER: Error-Resilient Spintronic Computational Random Access Memory for Scalable In-Memory Computation
}
\titlenote{Author preprint. Accepted to the Great Lakes Symposium on VLSI 2026 (GLSVLSI '26), June 22--24, 2026, Canandaigua, NY, USA. The Version of Record is available in the ACM Digital Library. DOI: 10.1145/3787109.3815240.}

\author{Sohan Salahuddin Mugdho}
\orcid{0009-0004-2670-285X}
\affiliation{%
  \department{Electrical and Computer Engineering}
  \institution{Iowa State University of Science and Technology}
  \city{Ames}
  \state{IA}
  \postcode{50010}
  \country{USA}
}

\author{Md. Shahedul Hasan}
\orcid{0000-0002-7182-0677}
\affiliation{%
  \department{Electrical and Computer Engineering}
  \institution{Iowa State University of Science and Technology}
  \city{Ames}
  \state{IA}
  \postcode{50010}
  \country{USA}
}

\author{Brahmdutta Dixit}
\orcid{0009-0008-4471-3877}
\affiliation{%
  \department{Electrical and Computer Engineering}
  \institution{University of Minnesota Twin Cities}
  \city{Minneapolis}
  \state{MN}
  \postcode{55455}
  \country{USA}
}

\author{Yang Lv}
\orcid{0000-0001-9062-309X}
\affiliation{%
  \department{Electrical and Computer Engineering}
  \institution{University of Minnesota Twin Cities}
  \city{Minneapolis}
  \state{MN}
  \postcode{55455}
  \country{USA}
}

\author{Jian-Ping Wang}
\orcid{0000-0003-2815-6624}
\affiliation{%
  \department{Electrical and Computer Engineering}
  \institution{University of Minnesota Twin Cities}
  \city{Minneapolis}
  \state{MN}
  \postcode{55455}
  \country{USA}
}

\author{Cheng Wang}
\authornote{Corresponding author: chengw@iastate.edu}
\orcid{0000-0002-8815-2871}
\affiliation{%
  \department{Electrical and Computer Engineering}
  \institution{Iowa State University of Science and Technology}
  \city{Ames}
  \state{IA}
  \postcode{50010}
  \country{USA}
}

\begin{abstract}
Deep neural networks (DNNs) have achieved state-of-the-art performance across diverse domains. However, typical Von Neumann compute paradigms face severe memory bottlenecks. Emerging near-memory and compute-in-memory approaches alleviate this but incur significant peripheral overhead. Computational Random Access Memory (CRAM) based on MRAM enables in-situ logic without peripheral overhead, offering a dense, energy-efficient solution. However, probabilistic MRAM switching induces gate-level errors that limit the scalability and reliability of CRAM for accelerating DNN. Moreover, the large number of sequential MRAM writes severely constrains CRAM throughput. To address these challenges, we propose an error-resilient CRAM (CRAM-ER) architecture for scalable in-memory matrix-vector multiplications (MVMs). Our error-aware hardware-software co-design framework leverages a hybrid spintronic-CRAM + CMOS adder-tree architecture to mitigate the impact of device-level errors, demonstrating MVM functionality with high area and energy efficiency. We further develop an error-aware model fine-tuning and fine-grained error correction for enhanced error resilience. Evaluations of the CMOS+spintronic hybrid architecture on DNN benchmarks show near-lossless accuracy while reducing CRAM latency by up to 2 orders of magnitude, outperforming CPU/GPU+high-bandwidth DRAM in both energy efficiency and energy-delay product.
\vspace{-10pt}
\end{abstract}


\maketitle
\vspace{-10pt}
\section{Introduction}
Deep Neural Networks (DNNs) have driven major advances in Artificial Intelligence (AI) across domains such as computer vision, medical diagnosis, autonomous vehicles, and natural language processing \cite{brief-history-of-ai,deeplearning}. However, data-intensive DNN workloads require heavy data movement between memory and compute units, causing the Von Neumann memory wall bottleneck \cite{case-for-emerging-memory}. Architectures that mitigate this include ``near-memory'' and ``in-memory'' computing systems \cite{paper12,issac, sram-cim}. 
Analog in-memory computing offers massive parallelism and high efficiency, but scalability is limited by analog non-idealities and peripheral overhead, especially analog-to-digital converters (ADCs). Digital in-memory computing performs multiply-and-accumulate (MAC) operations within the memory bank, but additional multipliers and adder trees incur large area overhead.
\begin{figure}[ht!]
    \centering
    \vspace{-15pt}
    \includegraphics[width=0.95\linewidth]{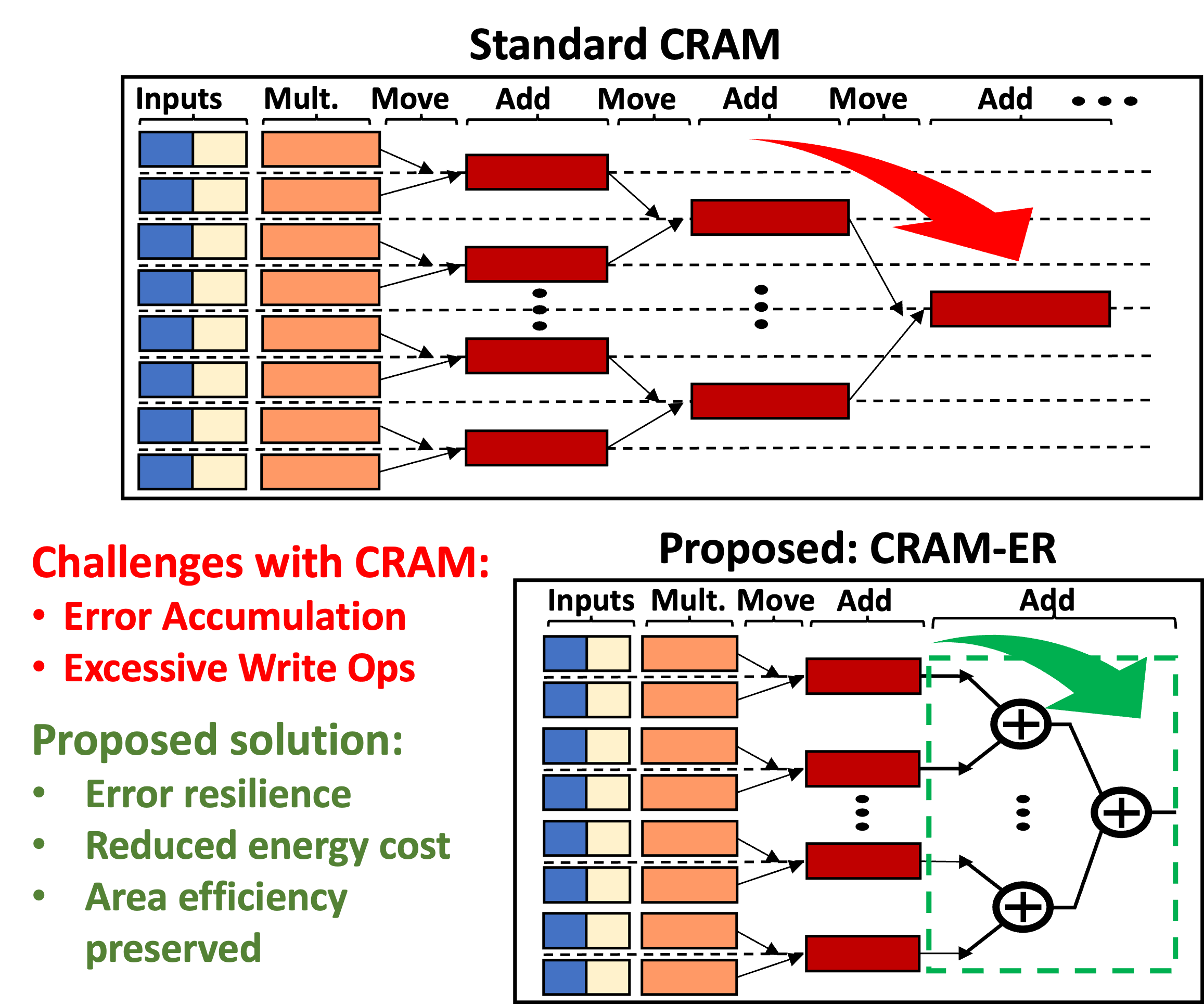}
    \vspace{-10pt}
    \caption{Overview of the challenges of standard CRAM (top), and the advantage of the proposed CRAM-ER design (bottom)}
    \label{fig:overview}
    \vspace{-10pt}
\end{figure}
Another processing-in-memory architecture is Computational Random Access Memory (CRAM), which executes logic operations directly in memory cells. With minimal peripheral overhead, CRAM performs \textit{in-situ} updates of results within the array. Dynamic random access memory (DRAM), spin transfer torque magnetic RAM (STT-MRAM), and Spin-Orbit Torque/Spin-Hall-Effect MRAM (SOT/SHE-MRAM) have all been evaluated for CRAM \cite{paper15, paper40, SHE-CRAM}. Among these, STT-MRAM is especially promising, offering non-volatility, non-destructive reads, high density, sub-10 ns access time, and write endurance above $10^{15}$ \cite{memory-table}. 

Although spintronic CRAM with demonstrated universal Boolean computation can in principle support arbitrary computation, prior work has focused only on small-scale, narrow tasks, such as integer arithmetic and logic operations \cite{paper12}, Support Vector Machine (SVM), and Genetic Algorithms \cite{paper16}.
It remains unexplored whether CRAM can scale to large, general DNNs beyond Binarized Multi-layer perceptrons \cite{paper15}\cite{paper14}. Moreover, the impact of technology-dependent, non-ideal device behaviors on the correctness and efficiency of application-level computation (e.g., multi-bit MACs) remains unclear.

In this work, we develop CRAM-ER, a scalable, error-resilient CRAM architecture for large-scale DNN workloads.  
We demonstrate the feasibility of CRAM and identify key challenges in implementing CRAM for multi-bit DNN workloads, including convolutional neural networks (CNN) and Vision Transformers (ViT) \cite{vit}. To the best of our knowledge, \textbf{our work is the first demonstration of CRAM for accelerating multi-bit MACs towards scalable DNN computation}. Our main contributions are:
\vspace{-3pt}
\begin{itemize}
    \item \textbf{We identify two key challenges for DNN acceleration with spintronic CRAM.} First, probabilistic MRAM switching causes severe accuracy loss due to accumulated errors in the dominating MAC operations. Second, high MRAM write overhead limits CRAM throughput and system efficiency. We address these issues with a spintronic+CMOS hybrid architecture that enables parallel in-situ multiplications and error-resilient additions. Partitioning MACs between CMOS and MRAM at the bit level provides an optimal trade-off between area overhead and processing efficiency.
    \item \textbf{We develop an adaptive hardware-software co-design framework} to mitigate impact of device-level errors in DNN inference. At the algorithm level, we model device- and gate-level error behavior with high fidelity and integrate it into a bit-error-aware DNN fine-tuning model. This is combined with a low-overhead partial error-correction (EC) circuit and an error-aware bit-level partitioning scheme to reliably handle error-prone multi-bit additions.
    \item \textbf{A comprehensive simulation framework} is developed to jointly evaluate inference accuracy and hardware efficiency of hybrid CRAM-ER. Running DNN benchmarks on CRAM-ER shows near-lossless accuracy with 10x better energy efficiency and 2x energy-delay product (EDP) improvement over the A100 GPU. Based on the insight that \textbf{memory write efficiency becomes the primary performance bottleneck}, we also evaluate a projected MRAM configuration with efficient writes based on the recent advances in devices, showing 16x EDP improvement over CPU/GPU references.
\end{itemize}
The rest of the paper is organized as follows. Sec.~\ref{sec:background} introduces basics on STT-MRAM and CRAM. Sec.~\ref{sec:related-works} reviews related compute-in-memory architectures.
Sec.~\ref{sec:evaluation} details the proposed CRAM architecture and the co-design approach.
Sec.~\ref{sec:evaluation} outlines the setup for software/hardware evaluation and Sec.~\ref{sec:results} presents the results. Sec.~\ref{sec:conclusion} concludes the paper.
\vspace{-5pt}
\section{Background}
\label{sec:background}
\subsection{STT-MRAM basics}
\label{subsec:sttmram}
STT-MRAM arrays use magnetic tunnel junctions (MTJs) with two magnetic layers (fixed and free) separated by an insulator. The resistance corresponding to their relative magnetic orientation (parallel (P) or anti-parallel (AP)) sets the tunneling magnetoresistance ratio (TMR), $TMR = \frac{R_{AP}-R_P}{R_P}$. Logic in CRAM based on STT-MRAMs relies on inherently probabilistic STT switching, with switching probability $P_{SW}$ described by the thermal activation model \cite{thermal-model}.
\begin{equation}
\label{eqn:psw}
P_{SW} = 1 - \exp\left(-\frac{t_p}{\tau}\right); \tau = \tau_0 \exp\left( \Delta \left[ 1 - \frac{V_p}{V_{c0}} \right] \right).
\end{equation}
Here, $\tau_p$ is the voltage pulse duration, $\tau$ the thermal activation lifetime, $\tau_0$ the attempt time, $\Delta$ the thermal stability factor, $V_p$ the voltage pulse amplitude, and $V_{c0}$ the intrinsic switching voltage. Table~\ref{tab:mtj} lists the MTJ device parameters used in this paper. CRAM logic operations based on these parameters are detailed in the next section.
\begin{table}[ht]
\centering
\vspace{-8pt}
\caption{STT-MRAM specifications based on \cite{paper13}. The marked (*) parameters are based on recent advances shown in \cite{adv-mram-data}}
\vspace{-5pt}
\label{tab:mtj}
\begin{tabular}{|l |l|}
\hline
\textbf{Parameters} & \textbf{Specifications} \\
\hline
MTJ type & STT-PMTJ (Heating+SOT *) \\
Material system & CoFeB/MgO/CoFeB (Ta/GdFeCo/Pt *) \\
MTJ diameter (nm) & 45 (20 *)  \\
TMR (\%) & 133 \\
RA product ($\Omega~\mu$m$^2$)  & 5 \\
$J_{c0}$ (A/cm$^2$) & $3.1 \times 10^6$ ($7 \times 10^8$ *)  \\
$t_{wr}$ (ns) & 3 (0.01 *) \\
$R_P$ (k$\Omega$) & 3.15 \\
$\Delta$ & 47 \\
$V_{c0}$ (V) & 0.155\\
\hline
\end{tabular}
\vspace{-15pt}
\end{table}
\begin{figure*}[h!]
    \centering
    \includegraphics[width=0.97\linewidth]{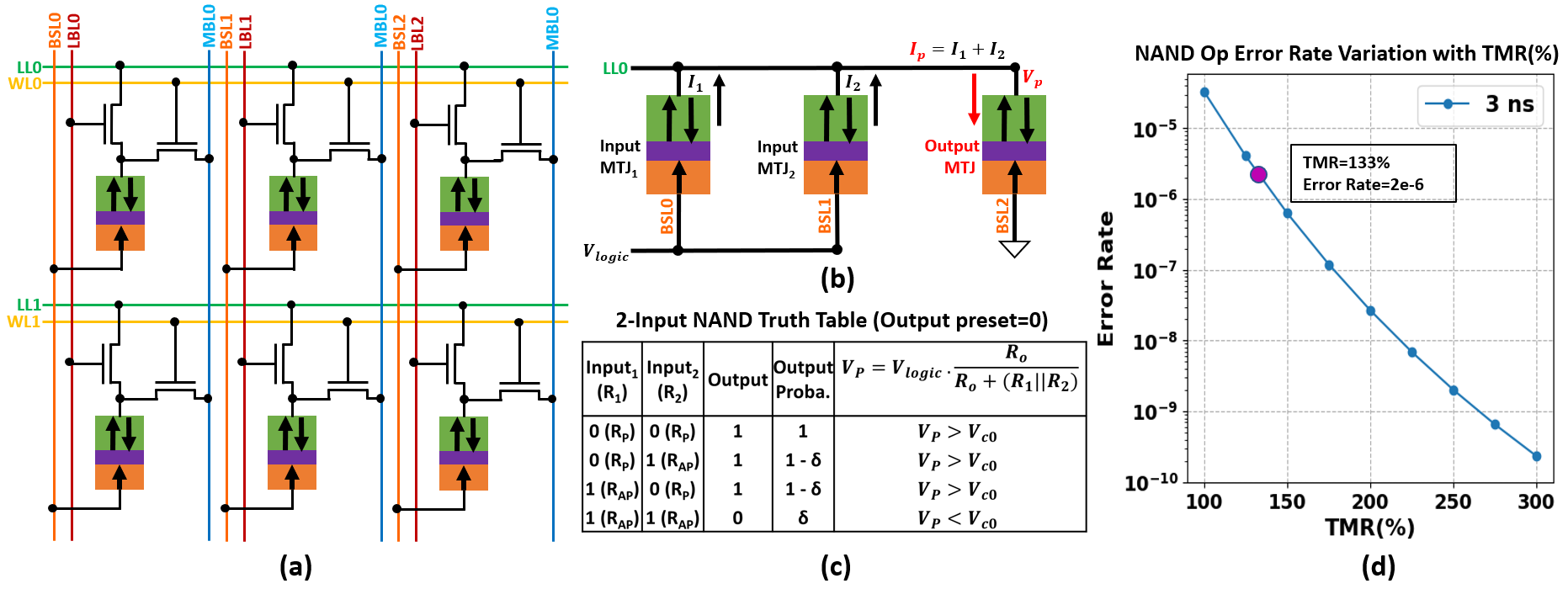}
    \vspace{-12pt}
    \caption{Working principle of a CRAM (a) CRAM array for logic in-memory, (b) 2-input logic operation through MTJ write, (c) probabilistic truth table of a NAND gate and (d) NAND gate-level error rate with varying TMR(\%) at 3ns write pulse.}
    \label{fig:background}
    \vspace{-15pt}
\end{figure*}
\subsection{Working principles of CRAM}
A typical CRAM slightly modifies the 1 Transistor 1 MTJ (1T1M) architecture \cite{paper57}, where an MTJ and an access transistor are in series between the bit select line (BSL) and the memory bit line (MBL), with the word line (WL) driving the transistor gate. 
CRAM adds a second transistor that connects the device to the logic line (LL) via the logic bit line (LBL), forming a 2T1M structure, as shown in Fig.~\ref{fig:background}(a), which enables logic operations. MTJ cells mapping the input operands and the output share an LL \cite{paper40}. A logic voltage $V_{logic}$ is then applied to the BSLs of the input MTJs while the BSL of the output MTJ is grounded, as in the 2-input, 1-output example in Fig.~\ref{fig:background}(b). 
When $V_{logic}$ is applied, the input MTJs generate currents based on their resistance levels whose sum flows through the output MTJ, creating a write voltage pulse $V_p$ across it. Depending on the input states, the output MTJ switches with probability $P_{SW}$, given by Eqn.~\ref{eqn:psw}.

Fig.~\ref{fig:background}(c) shows the truth table of a 2-input, 1-output NAND operation based on the previously described mechanism. CRAM supports all basic bitwise logic operations, including AND, OR, NAND, NOR, and MAJ \cite{paper12}. Because input operands and intermediate results remain entirely within the memory array, CRAM enables true in-memory digital operations \cite{paper12}. Notably, CRAM can use various memory devices; here, we evaluate STT-MRAM for its high density, CMOS compatibility, and strong endurance.
\vspace{-5pt}
\subsection{Current challenges with CRAM}
Despite its potential, CRAM faces scalability limits due to technology-dependent probabilistic write behavior. Each truth-table entry can have an output error rate $\delta$ ($0 \le \delta \le 1$). With proper tuning of $V_{logic}$, the NAND '00' input state can be ensured to have negligible errors, while '01', '10', and '11' each have error rate $\delta$ (see the `probabilistic output' column in Fig. 2(c)). 
In STT-MRAM CRAM, current-driven switching errors depend on the resistance levels and thus the TMR of the input MTJs. 
As shown in Fig.~\ref{fig:background}(d), a larger contrast between $R_P$ and $R_{AP}$ better separates the three distinct NAND input combinations (with '01' and '10' equivalent), reducing $\delta$ as TMR increases. Current STT-MRAM TMR is about 100–200$\%$; we use TMR = 133$\%$ as a conservative case, noting that higher TMR further improves CRAM functional performance. On the other hand, CRAM relies heavily on numerous MRAM write operations, so its efficiency strongly depends on MRAM write efficiency. Faster, lower-power MRAM writes can significantly improve CRAM performance.
\vspace{-5pt}
\section{Related Works}
\label{sec:related-works}
Recent compute-in-memory (CiM) and processing-in-memory (PiM) architectures have made progress in mitigating the von Neumann bottleneck, but key challenges in scalability, precision, and reliability remain. Analog CiM designs such as ISAAC \cite{issac}, PipeLayer \cite{rv6}, and the RRAM-based NeuRRAM chip \cite{rv1} use crossbar arrays for in-situ analog MACs, achieving high throughput and energy efficiency, but face peripheral overhead (DAC/ADC energy), computational inaccuracies due to, limited precision, IR-drop, and conductance variability \cite{fairxbar-paper,multibit-mram-paper}. Digital SRAM-based CiM designs like Neural Cache \cite{rv3}, charge and current-based analog SRAM-CiMs and time-domain or hybrid mixed-signal CiMs, support bit-wise or mixed-precision operations within memory arrays \cite{sramrv}, yet are limited by circuit complexity, low precision ($\le 8$ bits), and poor scalability to large networks \cite{sramrv}. DRAM-based PiM systems, including Ambit \cite{rv2}, DRISA \cite{rv4}, and CIDAN-3D \cite{rv14}, enable in-situ Boolean and arithmetic operations via multi-row activation or neuron-based processing elements, but suffer from data overwrites, high latency, and limited arithmetic flexibility.

In parallel, NVM-based PiM and CRAM architectures—such as GraphS \cite{rv7}, CRAFFT \cite{paper6}, SC-CRAM \cite{paper2}, spintronic CAM \cite{paper21}, and MOUSE \cite{paper14}—have shown promise for logic computation, FFT, matrix multiplication, and small-scale ML/BNN workloads, but remain confined to low-complexity tasks because probabilistic device switching and accumulated logic errors prevent reliable multi-bit DNN scaling. Reviews of SRAM-based CiM \cite{sramrv} and PiM-based DNN accelerators \cite{pimrv} compare these designs and highlight trade-offs among analog accuracy, digital scalability, and energy efficiency. In contrast, we target scaling STT-MRAM-based CRAM to complex multi-bit DNNs, mitigating switching-induced error accumulation via error-aware fine-tuning, partial error correction, and a CRAM–adder-tree hybrid architecture that improves functional accuracy while preserving CRAM’s intrinsic energy and density benefits.
\vspace{-15pt}
\section{Evaluation Methodology}
\subsection{Matrix-vector multiplication in CRAM} 
\label{sec:evaluation}
\begin{figure*}[h!]
    \centering
    \includegraphics[width=0.99\linewidth]{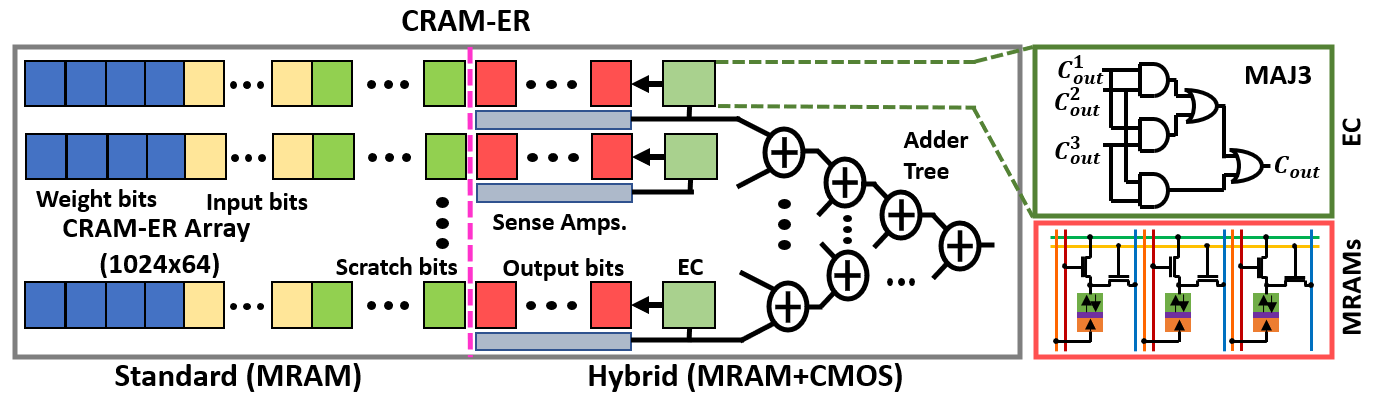}
    \vspace{-10pt}
    \caption{Detailed architecture of the CRAM-ER macro with low-overhead error correction (EC) mechanism and CMOS adder tree (left). A 1024x64 array offers a large number of rows for parallel operations and enough bits in each row to perform complex multi-bit operations (center). }
    \label{fig:arch}
    \vspace{-10pt}
\end{figure*}
In this work, we build on the all-NAND full adders (FAs) to hierarchically build ripple-carry adders and systolic multipliers \cite{paper12} for multi-bit addition and multiplication. 
Using these primitives, we design a dot-product engine, representative of matrix-vector multiplication (MVM). 
Both convolution and fully connected layers are expressed as sequences of MVMs. 
For each dot product, we perform element-wise multiplications with an array multiplier and then accumulate the partial products with a tree adder. Although the primitives support signed/unsigned integer and floating-point operations, we restrict our study to unsigned integers for simplicity.

The number of NAND instances grows with computation complexity. For example, a single all-NAND full adder (FA) requires 9 NAND operations to generate sum and carry bits. Consequently, a MAC between two 4-bit values using an array multiplier and tree adder requires over 200 sequential NAND operations, leading to high energy consumption and delay. In standard CRAM, this increased operation count, combined with device-level write errors, causes severe accuracy loss for complex multi-bit DNNs. In the next section, we present the hybrid CRAM-ER architecture, which enables efficient, error-resilient scaling for large DNN workloads.
\vspace{-5pt}
\subsection{Efficient Error-resilient hardware design}
To mitigate error accumulation and accuracy loss in STT-MRAM CRAM, we propose an error-resilient CRAM (CRAM-ER) framework that combines a low-overhead partial error-correction scheme with a hybrid CRAM-adder tree architecture. This scheme eliminates many write operations during accumulation, enabling efficient MVM. Paired with an error-aware algorithm-level fine-tuning strategy, the framework achieves near-lossless DNN inference accuracy.

\textbf{Selective Bit Error Correction.} Fig.~\ref{fig:arch} shows the CRAM-ER architecture. Built on a typical CRAM macro \cite{paper14}, CRAM-ER adds low-overhead error correction (EC) circuits and an adder tree. The EC mechanism is a MAJ3 circuit that takes three output bits from the same operation and writes their majority as the final output. Because in-array sense amplifiers are required to read STT-MRAM cells, they are attached only to a subset of cells in each row. This allows the EC circuit and adder tree to directly access output bits while limiting area overhead.

Applying EC incurs notable energy and latency overhead for each output. Using recent results in error-aware heterogeneous memory design \cite{mram-ber-paper}, we observed that the most significant carry bits are most critical for accuracy. Based on this insight, we optimized the EC scheme: in CRAM-ER, EC is applied at bit-level granularity, and only the \textbf{final carry} bit is generated three times to apply correction at each level of the adder tree. 

\textbf{Hybrid Design with Partial CMOS Adder Trees.} Despite error correction, greater accumulation in CRAM increases gate-level errors, and relying solely on EC incurs high energy and latency. This motivates our hybrid design with CMOS adder trees. Inspired by error-free digital CiM, we partition vectorized MAC operations between MRAM-based CRAM and CMOS. However, full-precision digital adder trees cause large area overheads. To balance CRAM’s area efficiency with error mitigation, we vary the MAC partitioning between CRAM and CMOS adders. In dot-product operations, we find that accumulating a large fraction (75$\%$) of products in CRAM has limited impact on overall accuracy. Thus, CRAM-ER performs all element-wise multiplications and most accumulations in CRAM, while a smaller digital adder tree completes the remaining accumulations error-free. This partial adder tree also removes many MRAM write operations, greatly improving efficiency while keeping area overhead low.

\textbf{Simulation Development}. We develop a Python-based framework for evaluating system-level performance of CRAM-ER based on the MTJ specification in Table~\ref{tab:mtj}. The 2T1M memory cell is designed based on the GlobalFoundries 22nm Fully Depleted Silicon on Insulator (FDSOI) Process Design Kit (PDK) \cite{GF22FDX}. The data for the peripheral components is sourced from NVsim ~\cite{nvsim} based on the 22nm technology. The EC circuit and the adder tree are first synthesized using Cadence’s RTL Compiler tool, Genus, with the 45nm NCSU PDK library \cite{ncsuFreePDK45}, then scaled down to 22nm technology node following DeepScale \cite{sarangi2021deepscaletool}.
\vspace{-10pt}
\subsection{Error modeling and error-aware fune-tuning} 
We develop a Pytorch-based framework to simulate the impact of DNN processing in CRAM and CRAM-ER. We first implement the probabilistic NAND gate described in Sec.~\ref{fig:background}, then implement the all-NAND FA, ripple carry adder, systolic array multiplier, and scale the simulator to perform full-scale matrix multiplications to process DNNs. However, simulating the entire DNN model at the gate level granularity is extremely slow and not feasible for large models. Thus, we utilize our CRAM simulation framework to estimate the bit-level error rates of performing matrix multiplication with a fixed array size, and perform large-scale experimentation based on the estimated error rates. Furthermore, to improve the DNN model accuracy beyond the hardware mitigation strategies, we develop an error-aware fine-tuning scheme. Our error-aware fine-tuning process follows the general steps described below:
\begin{enumerate}
    \item Train the baseline DNN model with Q-bit quantization;
    \item Represent trained weights and inputs as Q-bit unsigned integer numbers;
    \item Simulate a large number of matrix multiplications of a fixed array size and gate level error rate using probabilistic all-NAND primitives;
    \item Estimate bit-level error rate at the output;
    \item Fine-tune DNN model with error or perform inference with error.
\end{enumerate}
\vspace{-10pt}
\begin{table}[ht]
\centering
\scriptsize
\setlength{\tabcolsep}{4pt}
\renewcommand{\arraystretch}{1.1}
\caption{Algorithmic accuracy evaluation of the Q-baseline, CRAM only, CRAM+EC, and CRAM+EC+Finetune configurations.}
\label{tab:combined-acc}
\vspace{-15pt}
\begin{subtable}{\linewidth}
\centering
\caption{LeNet-5 on MNIST Dataset}
\vspace{-5pt}
\label{tab:lenet5-acc}
\resizebox{0.9\textwidth}{!}{%
\begin{tabular}{|c|c|l|c|}
\hline
\textbf{Error Rate} & \textbf{Adder Tree (\%)} & \textbf{Configuration} & \textbf{Accuracy (\%)} \\
\hline

\multirow{12}{*}{\begin{tabular}{c}
2e-6 \\
TMR=133\%
\end{tabular}}

& - & Q-baseline & \textbf{98.65} \\ \cline{2-4}

& \multirow{2}{*}{0} 
& CRAM only & 25.50 \\
& & CRAM+EC & 32.75 \\ \cline{2-4}

& \multirow{3}{*}{12.5} 
& CRAM only & 64.80 \\
& & CRAM+EC & 88.74 \\
& & CRAM+EC+Finetune & 96.41 \\ \cline{2-4}

& \multirow{3}{*}{25} 
& CRAM only & 91.48 \\
& & CRAM+EC & 95.03 \\
& & CRAM+EC+Finetune & \textbf{98.26} \\ \cline{2-4}

& \multirow{3}{*}{50} 
& CRAM only & 98.03 \\
& & CRAM+EC & 98.15 \\
& & CRAM+EC+Finetune & \textbf{98.64} \\ \cline{2-4}

\hline
\end{tabular}
}
\end{subtable}

\vspace{10pt}
\begin{subtable}{\linewidth}
\centering
\vspace{-8pt}
\caption{ResNet-20 on CIFAR-10 Dataset}
\vspace{-5pt}
\label{tab:resnet20-acc}
\resizebox{0.9\textwidth}{!}{%
\begin{tabular}{|c|c|l|c|}
\hline
\textbf{Error Rate} & \textbf{Adder Tree (\%)} & \textbf{Configuration} & \textbf{Accuracy (\%)} \\
\hline

\multirow{16}{*}{\begin{tabular}{c}
2e-6 \\
TMR=133\%
\end{tabular}}

& - & Q-baseline & \textbf{89.77} \\ \cline{2-4}

& \multirow{3}{*}{0} 
& CRAM only & 14.62 \\
& & CRAM+EC & 23.54 \\
& & CRAM+EC+Finetune & 52.12 \\ \cline{2-4}

& \multirow{3}{*}{12.5} 
& CRAM only & 41.46 \\
& & CRAM+EC & 46.92 \\
& & CRAM+EC+Finetune & 78.83 \\ \cline{2-4}

& \multirow{3}{*}{25} 
& CRAM only & 52.54 \\
& & CRAM+EC & 80.00 \\
& & CRAM+EC+Finetune & \textbf{89.08} \\ \cline{2-4}

& \multirow{3}{*}{50} 
& CRAM only & 68.21 \\
& & CRAM+EC & 84.00 \\
& & CRAM+EC+Finetune & \textbf{89.17} \\ \cline{2-4}

& \multirow{3}{*}{100} 
& CRAM only & 88.54 \\
& & CRAM+EC & 88.58 \\
& & CRAM+EC+Finetune & \textbf{89.21} \\ \hline

\end{tabular}
}
\end{subtable}
\vspace{-15pt}
\end{table}
\section{Results}
\label{sec:results}
\subsection{Inference accuracy evaluation} 

We perform the algorithmic accuracy on the MNIST dataset \cite{mnist} using a slightly modified version of the LeNet-5 model \cite{lenet}, and on the CIFAR-10 dataset \cite{cifar10} using the ResNet-20 model \cite{resnet}, representing a simple and a more complex workload, respectively. 
Tables~\ref{tab:combined-acc}(a) and ~\ref{tab:combined-acc}(b) demonstrate the algorithmic performance of the LeNet-5 model on the MNIST dataset and the ResNet-20 model on the CIFAR-10 dataset, respectively.
For both workloads, we prepare our baseline as a \textbf{4b input and 4b weight} quantized model referred to as the \textbf{Q-baseline}. We compare the inference accuracy of different CRAM-based configurations for both workloads. 
The \textbf{CRAM only} configuration processes all MVM operations entirely on CRAM without any error correction or fine-tuning. Error mitigation techniques, ``EC" and ``Finetune" are added subsequently.
Various hybrid configurations are evaluated with different percentages of the ADD operations partitioned to the CMOS adder tree circuits. 
``Adder Tree (\%)'' indicates the percentage of the multi-bit ADD operations done in CMOS. 
Our results of the \textbf{CRAM+EC} and the \textbf{CRAM+EC+Finetune} configurations demonstrate significant accuracy improvement over the \textbf{CRAM only} version. Furthermore, by performing as low as 25\% of the accumulations on the digital adder tree, our most optimized configuration \textbf{CRAM+EC+Finetune} shows near-lossless performance compared to the \textbf{Q-baseline}.

\begin{figure}[h!]
    \centering
    \includegraphics[width=0.99\linewidth]{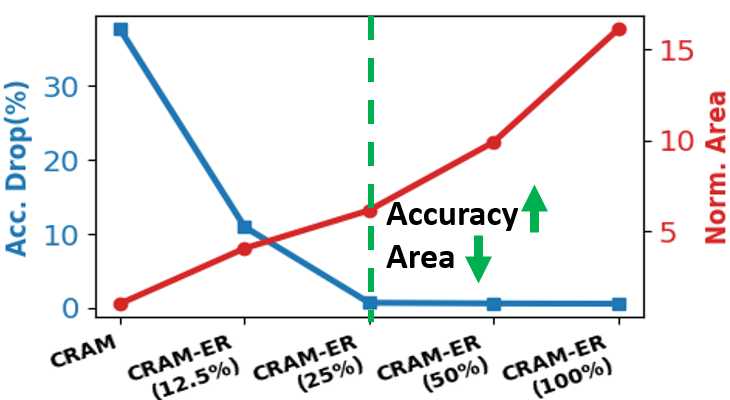}
    \vspace{-5pt}
    \caption{Accuracy Drop (\%) and Normalized Area vs different CRAM designs. The configurations include a typical CRAM design and CRAM-ER with different adder tree sizes.}
    \vspace{-15pt}
    \label{fig:acc-area}
\end{figure}

\textbf{Accuracy-Area Trade-off.} Fig.~\ref{fig:acc-area} shows the accuracy-area trade-off of CRAM-ER. We evaluate the accuracy drop (\%) of our most optimized \textbf{CRAM+EC+Finetune} configuration on the CIFAR-10 dataset using the ResNet-20 model, based entirely on CRAM, or the increasing percentage of digital adder tree accumulation given by the \textbf{CRAM-ER (X\%)} design, where (X\%) is the percentage of accumulations on the adder tree.
The standard CRAM suffers from significant accuracy loss, while near-lossless algorithmic performance can be achieved by performing as few as 25$\%$ of the accumulations on the adder tree, i.e., the CRAM-ER(25\%) design. Normalized Area shows that pushing more accumulations into the adder tree beyond the near-lossless accuracy point results in undesired area overhead. In contrast, further pushing accumulations into CRAM will improve area efficiency, but at the cost of reduced accuracy.
\vspace{-10pt}
\begin{figure}[h!]
    \centering
    \includegraphics[width=\linewidth]{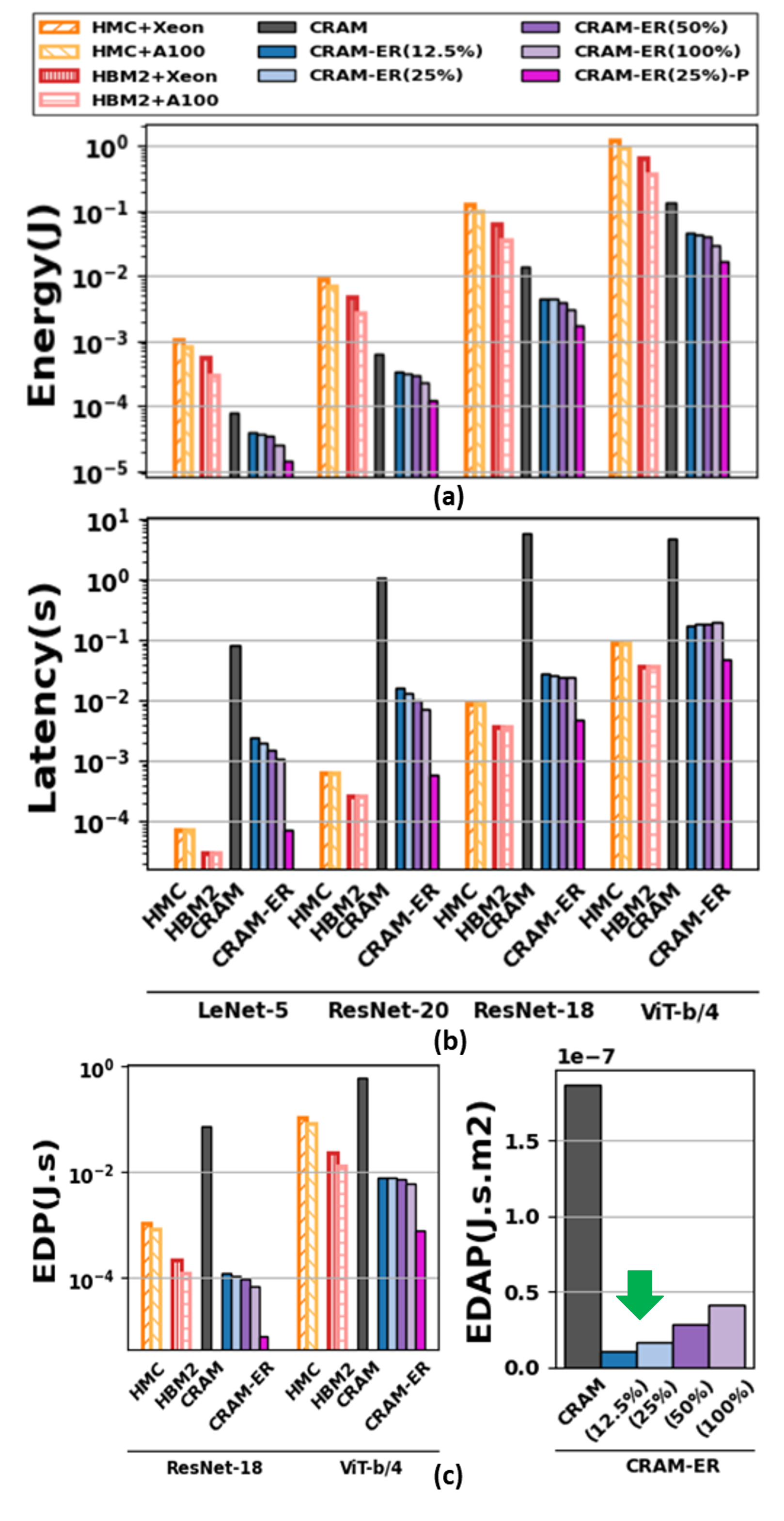}
    \vspace{-20pt}
    \caption{System-level performance evaluation of NMC platforms, a typical CRAM, and the proposed CRAM-ER architectures on different DNN workloads. (a) shows the total energy $(J)$, (b) shows the total latency $(s)$ for a single inference. (c) shows the energy-delay product (EDP)$(J)$ of the ResNet-18 and ViT/b-4 workloads, and energy-delay-area product (EDAP)$(J.s.mm^2)$ of a CRAM and the proposed CRAM-ER(X\%) designs on the ViT/b-4 model.}

    \label{fig:energy-lat}
    \vspace{-20pt}
\end{figure}
\subsection{System-level hardware performance} 
We illustrate the advantage of CRAM at removing the memory bottleneck by comparing the performance of CRAM with standard GPU/CPU+DRAM paradigms, where the DRAM here considers recent main memory technology advances such as Hybrid Memory Cube (HMC) \cite{hmc}, and High Bandwidth Memories (HBM2) \cite{hbm2}. Fig.~\ref{fig:energy-lat} shows the system-level energy and latency of the CRAM and CRAM-ER architectures compared to near-memory computing (NMC) platforms based on different high bandwidth memory systems, including HMC and HBM2. We use the per-MAC energy of the Intel Xeon Platinum 8480 CPU \cite{intel_xeon_specs,intel_xeon_flops,intel_xeon_int}, and the Nvidia A100 GPU \cite{gpu_a100_mac}, as the compute systems based on their respective performance datasheets, and appropriately scale the MAC energy to 22nm technology and 4b operation. We conservatively assume that these computing systems are optimized, so the latency is dominated by memory bandwidth. 

Based on these memory technologies and computing platforms, we evaluate four configurations \{\textbf{HMC}, \textbf{HBM2}\}+\{\textbf{Xeon}, \textbf{A100}\} for comparison. We evaluated CRAM-based configurations, including a conventional \textbf{CRAM} and the \textbf{CRAM-ER(X\%)} architectures as shown in Fig.~\ref{fig:acc-area}. Here, the (X\%)=\{12.5\%, 25\%, 50\%, 100\%\} represent the percentages of accumulation performed through the adder tree. 
Fig.~\ref{fig:energy-lat} summarizes the (a) energy, (b) latency, and (c) the energy-delay product (EDP) of the configurations evaluated in four representative DNNs: LeNet-5 on the MNIST data set and ResNet-20, ResNet-18, ViT-b/4 on the CIFAR-10 data set. Importantly, the primary advantage of CRAM-ER is energy efficiency. As shown in  Fig.~\ref{fig:energy-lat}(a), the proposed hybrid CRAM-ER exhibits further reduction of energy consumption compared to the CRAM baseline, achieving up to 40x and 10x improvements over \textbf{HMC + Xeon CPU} and \textbf{HBM2 + A100 GPU} respectively.

It is important to note that the CRAM baseline has higher latency than the CPU/GPU reference configurations, primarily due to the required large number of costly MRAM write operations for the accumulation in MAC. Such high latency becomes a bottleneck, preventing the adoption of MRAM-based CRAM for large-scale MAC computation. The hybrid CRAM-ER shows a remarkable 20-200x reduction in latency compared to conventional CRAM across all workloads, demonstrating the impact of the CMOS+spintronics hybrid approach in eliminating the latency bottleneck. 

In addition, to provide deeper insights into the technological potential of CRAM, we investigate how technological improvements in spintronic write performance may affect the performance of the CRAM-ER design. Based on a recently demonstrated device characteristics from \cite{adv-mram-data}, we obtained specifications for a write-efficient MRAM cell configuration and extended our analysis based on the specifications summarized in Table~\ref{tab:mtj}.
We evaluate a \textit{projected} CRAM-ER(25\%) configuration using the device with low-energy writes, labeled as \textbf{CRAM-ER(25\%)-P}. As shown in Fig.~\ref{fig:energy-lat}, CRAM-ER(25\%)-P consistently improves both energy efficiency and latency on top of CRAM-ER, achieving up to 70x energy efficiency compared to CPU and GPU while reaching near-HBM2 throughput.

\textbf{Efficiency-Area Trade-off}. The left panel of Fig.~\ref{fig:energy-lat}(c) shows the energy-delay product (EDP) of the NMC, and CRAM/CRAM-ER designs on the ResNet-18 and ViT-b/4 workloads. While the standard CRAM design underperforms EDP relative to NMC platforms by a large margin, our hybrid CRAM-ER designs significantly improve energy efficiency and close the latency gap, achieving up to 2x EDP improvement for larger workloads (16x for the projected CRAM-ER(25\%)-P). Furthermore, the energy-delay-area product (EDAP) of CRAM and hybrid CRAM-ER is shown in the right panel of Fig.~\ref{fig:energy-lat}(c). The co-designed hybrid CRAM-ER (25\%) achieves an optimal trade-off in the EDAP metric compared to the standard CRAM or other hybrid configurations, thanks to savings from reducing the large area overhead of adder trees while maintaining robust error resiliency.
\vspace{-10pt}
\section{Conclusion}
\label{sec:conclusion}
We present CRAM-ER, an error-resilient Spintronic Computational Random Access Memory architecture to enable scalable MRAM-based in-memory processing of multi-bit DNN workloads. By identifying probabilistic switching–induced error accumulation and massive MRAM writes as the primary barriers to scalability, we introduced a hybrid CMOS+spintronics architecture that handles gate-level errors while reducing write overheads through hardware-software co-designed approaches. 
Evaluations of representative DNNs demonstrate error resilience and substantial gains in system-level energy efficiency. Overall, CRAM-ER establishes a practical pathway toward dense, energy-efficient, and scalable CRAM-based DNN acceleration.
\vspace{-10pt}
\begin{acks}
This work is supported in part by National Science Foundation (NSF) Grant No. 2441290 and NSF Grant No. 2534279.
\end{acks}
\bibliographystyle{ACM-Reference-Format}
\setcitestyle{numbers}
\bibliography{ref}

\end{document}